\documentclass[letter]{aa} 
\usepackage{amsmath}
\usepackage{graphicx}
\usepackage{graphics}
\usepackage{subfigure}
\usepackage{txfonts}
\usepackage{natbib}
\bibpunct{(}{)}{;}{a}{}{,}

\def\ltsima{$\buildrel<\over\sim$}
\def\lsim{\lower.5ex\hbox{\ltsima}~}
\def\gtsima{$\buildrel>\over\sim$}
\def\gsim{\lower.5ex\hbox{\gtsima}~}

\def\teff{\ifmmode T_{\rm eff} \else $T_{\mathrm{eff}}$\fi}

\def\lya{Ly$\alpha$} 
\def\ha{H$\alpha$} 
\def\hb{H$\beta$}

\def\ebv{$E_{B-V}$}

\def\fesc{$f_{esc}$}

\def\cm2{cm$^{-2}$}

\def\ewlya{$EW_{\mathrm{Ly}\alpha}$}

\def\hi{H{\sc i}}

\def\nii{N{\sc ii}}

\def\nh{\ifmmode N_{\mathrm{HI}}\else $N_{\mathrm{HI}}$\fi}

\def\vexp{\ifmmode v_{\rm exp} \else v$_{\rm exp}$\fi}
\def\taua{\ifmmode \tau_{a}\else $\tau_{a}$\fi}

\begin{document}
%
 \title{Empirical estimate of \\
  \lya\ escape fraction in a statistical sample of \lya\ emitters}
\subtitle{}
\author{H. Atek\inst{1}, D. Kunth\inst{1}, D. Schaerer\inst{2,3}, M. Hayes\inst{2}, J. M. Deharveng\inst{4}, G. \"Ostlin\inst{5}, J. M. Mas-Hesse\inst{6}}

\institute{Institut d'Astrophysique de Paris (IAP), 98bis boulevard Arago, 75014 Paris, France
       \and Observatoire de Gen\`eve, Universit\'e de Gen\`eve, 51 Ch. des Maillettes, 1290 Sauverny, Switzerland
       \and
Laboratoire d'Astrophysique de Toulouse-Tarbes, 
Universit\'e de Toulouse, CNRS,
14 Avenue E. Belin,
31400 Toulouse, France
\and 
Laboratoire d'Astrophysique de Marseille, UMR 6110 CNRS/Universit\'e de Provence, 38 rue Joliot-Curie, 13388 Marseille Cedex 13, France
\and Oskar Klein Center for Cosmoparticle physics, Department of Astronomy, Stockholm University, 10691 Stockholm, Sweden
\and 
Centro de Astrobiolog\'{\i}a (CSIC-INTA), POB 78, E28691 Villanueva de la Ca\~nada, Madrid, Spain
}
\authorrunning{Atek et al}
\titlerunning{\lya\ escape fraction in a statistical sample of \lya\ emitters}

\date{Received date; accepted date}

\abstract{The Lyman-alpha (\lya) recombination line is a fundamental tool for galaxy evolution studies and modern observational cosmology. However, subsequent interpretations are still prone to a number of uncertainties. Besides numerical efforts, empirical data are urgently needed for a better understanding of the \lya\ escape process.}
{We empirically estimate the \lya\ escape fraction in a statistically significant sample of galaxies in a redshift range $z \sim 0 - 0.3$. This estimate will constrain interpretations of current high-redshift \lya\ observations.}
{An optical spectroscopic follow-up of a sub-sample of 24 \lya\ emitters detected by GALEX at $z \sim 0.2 - 0.3$, combined with a UV-optical sample of local starbursts, both with matched apertures, allow us to quantify the dust extinction through Balmer lines, and to estimate the \lya\ escape fraction from the \ha\ flux corrected for extinction in the framework of the recombination theory. }
{ The global escape fraction of \lya\ radiation spans a wide range of values and \fesc(\lya) 
clearly decreases with increasing nebular dust extinction E(B-V). 
Several objects show \fesc(\lya) greater than \fesc(continuum), which may be taken as observational evidence for a clumpy ISM geometry or for
an aspherical ISM. 
Selection biases and aperture size effects may still prevail between $z \sim 0.2 - 0.3$ Lyman-alpha emitters (LAEs) and local starbursts and may explain the difference observed for \fesc(\lya).}
{} 


\maketitle

\section{Introduction}

Considerable progress has been made in the last years in the detection and characterization of distant galaxy populations thanks, in particular, to 8-10m class telescopes with large field of view instruments. In this context, the \lya\ emission line is of particular interest, since it remains the brightest spectral signature of remote young galaxies \citep{pp67,schaerer03}. As a result, high redshift galaxies are now being routinely detected thanks to the Lyman Break selection and/or emission line surveys \citep[eg.][and references therein]{gronwall07,ouchi08,nilsson08}; this situation is likely to improve with upcoming Extremely Large Telescopes (ELTs) and the James Webb Space Telescope (JWST). A comparable survey is now available for the first time at low redshift \citep[$z \sim 0.2-0.35$,][]{deharveng08} thanks to the GALEX (Galaxy Evolution Explorer) UV capabilities.


The \lya\ line proves an invaluable tool in a cosmological context and is used in a wide variety of applications.
 \lya\ has been used in recent studies to probe early stages of galaxy formation, estimate the star formation rate, trace large scale structures, identify potential hosts of population III stars and place constraints on cosmic reionization at z \gsim\ 6. 
However, in order to ensure a proper interpretation of these very promising \lya-oriented studies, one first must establish a robust calibration of the many parameters that control the complex transport of this line. The determination of the amount of \lya\ radiation that escapes from the host galaxy is certainly the most important step toward understanding how various galaxy properties may distort the interpretation of \lya\ observations. 
Although the order of importance of these parameters has been extensively discussed \citep[e.g.][]{verhamme08,schaerer08,atek09a}, 
 empirical evidence based upon a large sample of galaxies is still missing.

From the International Ultraviolet Explorer (IUE) to the Hubble Space Telescope (HST) era, spectroscopic and imaging observations of nearby star forming galaxies have played a key role in identifying the main parameters responsible for the \lya\ escape in a given galaxy. Recent high resolution \lya\ imaging results clearly demonstrate the importance of resonant scattering, evidenced locally by very high \lya/\ha\ ratios and an outstanding large \lya\ scattering halo \citep[e.g.][]{hayes07,atek08,ostlin08}. However, most results so far have no statistical bearing and are still difficult to generalize, because the sample is not only small but consists of specific ``hand-picked'' objects. We propose here to improve this situation by using a larger sample of 24 \lya\ emitting galaxies at $0.2 \la z \la$ 0.35 found by GALEX. We have carried out a spectroscopic follow-up of a southern sub-sample with EFOSC2 on the ESO New Technology Telescope (NTT). This enables us to analyze how the \lya\ emission is related to many physical properties of galaxies. We also re-analyzed UV-optical spectra of 11 local starbursts. For the first time, these large aperture observations allow us to determine empirically the \lya\ escape fraction in a large sample of galaxies and to examine its dependence on dust extinction, if any.



\section{Observations }
\subsection{The GALEX sample}
\label{galex_sample_sec}
 96 \lya\ emitting galaxies at $z \sim 0.2 - 0.35$ were found by \citet{deharveng08} in the far ultraviolet  ($1350 \AA-1750 \AA$) from a {\it GALEX} slitless spectroscopic survey. 
Five fields covering a total area of 5.65 deg$^{2}$ were used to extract all continuum spectra with a minimum signal-to-noise ratio (S/N) per resolution element of 2 in the FUV. \lya\ emitters are then visually selected on the basis of a potential \lya\ emission feature, which naturally leads to a threshold of \ewlya\ $\gsim 10 \AA$. Data reduction and field characteristics are described in more detail in \citet{deharveng08}.

\subsection{Spectroscopic follow-up}
\label{follow-up_sec}

Spectroscopic observations of 24 of the 31 galaxies in the Chandra Deep Field South (CDFS) and ELAIS-S1 fields were performed
with {\it EFOSC2} on the {\it NTT} at ESO La Silla. Observational conditions were very good with photometric sky and sub-arcsec seeing ($0.5\arcsec - 1\arcsec$).
Two instrumental setups were used in long slit mode: (1) a spectrophotometric mode with a 5\arcsec\ slit, allowing observations to encompass the whole galaxy (20 out of 24 objects);
and (2) a spectroscopic mode with a 1\arcsec\ slit, giving a better spectral resolution enabling us to correct \ha\ data for \nii\ contamination. Both settings were used in combination with Grism \#13, covering a large wavelength range in the optical domain ($3690-9320 \AA$).
A binning of $2 \times 2$ is used and corresponds to a plate-scale of 0.24\arcsec\ px$^{-1}$ and a spectral resolution of FWHM $\sim$ 12 \AA\ (for 1\arcsec\ slit spectra). To avoid second order contamination that affects the longer wavelength range, an order sorting filter has been mounted to cut off light blue-ward of 4200 \AA. Observational settings and the mean exposure time per object are summarized in Table \ref{tab:observations}.

\begin{table}[htb]
\begin{center}
\centering
\begin{minipage}{\textwidth}
\caption{NTT EFOSC2 observations.}
\label{tab:observations}
\renewcommand{\footnoterule}{}  
\begin{tabular}{l l c c c}
\hline
\hline
\\
Mode                               &      Slit              &  Exptime (sec) & Nb Obj  & Grism\\
\hline \\
Spectroscopy                 &   1\arcsec        &   3000              &        24    & 13                \\
Spectrophotometry        &   5\arcsec        &  1800              &        20    & 13                 \\ 
\\
\hline  
\end{tabular}
\end{minipage}
\end{center}
\end{table}

The EFOSC2 spectra were reduced and calibrated using standard {\tt IRAF} routines. 
The aperture extraction of 1D spectra was performed through the {\tt DOSLIT} task. 
{Spectra were flux calibrated using a mean sensitivity function determined by observations of standard stars (Feige110, HILT600, LTT1020, EG21) from the \citet{oke90} catalog.}

\subsection{\it{IUE} starburst sample}
We have re-analyzed UV-optical spectra of 11 local starburst galaxies, presented in \citet{mcquade95} and \citet{storchi95} \citep[see also][]{giavalisco96} that are distant enough to separate the \lya\ feature of the galaxy from geocoronal \lya\ emission. In this way, we have complete control on the \lya\ emission measurement method. Indeed, the definition of a \lya\ emitter/absorber could be ambiguous for P Cygni profiles or an emission blended with absorption. Therefore, we consider here only the net \lya\ flux. Data consist of {\it IUE} UV ($1200 - 3300$ \AA) spectra combined with ground-based optical spectra with a matched aperture (20\arcsec\ $\times$ 10\arcsec).
The line measurements have been performed following the same procedure used for the {\it NTT} spectra. We included in our analysis only net \lya\ emitters, i.e. with \ewlya\ $> 0$. Furthermore, for common objects between {\it IUE} and the \citet{atek08} samples, we retained the latter to minimize aperture size effects (see below). We then obtained five {\it IUE} objects.



\subsection{Emission line measurements}
\label{emission_sec}

All spectra were analyzed using the {\tt SPLOT} package in {\tt IRAF}. The redshift was measured using the wavelength position of several lines and line measurements were performed interactively on rest-frame spectra. We confirm with a better accuracy the redshift determinations based on the blind search for \lya\ features in the GALEX survey.


Fluxes and equivalent widths (EWs) were measured for \ha, \hb\ and [\nii] 6548, 6584 \AA. For most spectra, the \ha\ line (6563 \AA) is blended with [\nii] lines, even for the 1\arcsec\ slit observations. In this case, a deblending routine is used within {\tt SPLOT} to measure individual fluxes in each line. Then, the \nii/\ha\ line ratio is used to correct the spectrophotometric observations for [\nii] contamination. It appears that the dust extinction (cf. Sect.\ \ref{sec:extinction}) is sensitive to the aperture size, since our 1\arcsec\ slit observations, targeting the center of the galaxies, led in general to a higher extinction. But this does not affect the \nii\ correction, which is relatively insensitive to dust extinction.
To correct for underlying stellar absorption in the Balmer lines, we assumed a constant equivalent width of 2 \AA, typical for
starburst galaxies \citep{tresse96, gonzalez99} .


To determine uncertainties in the line fluxes, we ran 1000 Monte Carlo simulations in which random Gaussian noise, based on the data noise, is added to a noise-free spectrum. Then, emission lines were fitted.
The computed MC errors depend essentially on the S/N quality of spectra. Error propagation is applied through the calculation of all the quantities described above and the line ratios, extinction etc, computed hereafter.       

Using BPT diagrams \citep{baldwin81,veilleux87}, and {\it Chandra} X-ray observations, we have identified at most three galaxies possibly excited by an active galactic nucleus (AGN), which represents up to 12.5 \% of our sample. Studying similar $z \sim 0.3$ samples, Scarlata et al. (2009, submitted) find a comparable value (17 \%), while  \citet{finkelstein09b} claim a much higher fraction (around 43 \%). We therefore removed these objects from our analysis. This diagnostic will be thoroughly addressed in Atek et al. (in prep).

\subsection{Extinction}
\label{sec:extinction}
Reddening along the galaxy line of sight is caused by interstellar dust extinction. The reddening contribution of our Galaxy is negligible for our objects. Then, the extinction coefficient, $C(H\beta)$, intrinsic to the observed object can be calculated using the Balmer ratio between \ha\ and \hb:
\begin{equation}
\label{c_equation}
\frac{f(H\alpha)}{f(H\beta)} = R \times 10^{- C [S(H\alpha) - S(H\beta)]}
\end{equation} 
where f(\ha) and f(\hb) are the measured integrated fluxes and $R$ is the intrinsic Balmer ratio. We use here a value of $R=2.86$, assuming case B recombination theory and a temperature of 10$^4$ K \citep{osterbrock89}.  
$S(H\alpha)$ and $S(H\beta)$ are determined from the \citet{cardelli89} extinction law. 
The colour excess \ebv\ is then simply computed using Eq. \ref{c_equation} and the relation $E(B-V) = C/1.47$. 
Due to photometric calibration errors and probably a stronger stellar absorption than assumed here, few objects have a negative extinction with large uncertainties. This could also be the result of enhanced \hb\ emission from a reflection nebula.

\section{\lya\ escape fraction}

\begin{figure}[htbp]
 \centering
  \includegraphics[width=9cm,height=7.5cm]{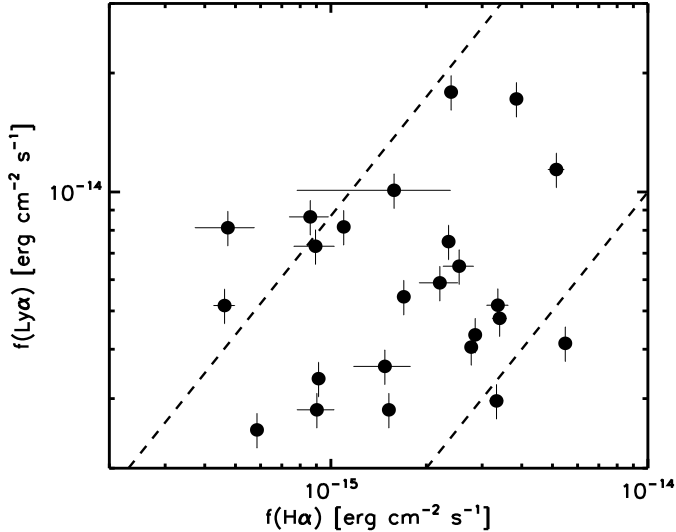}
 \caption{\lya\ versus \ha\ fluxes for the {\it GALEX} sample. The dashed line represent \lya/\ha=8.7 (case B) and \lya/\ha=1. The local objects (not represented here) show higher fluxes and lie in general close to the line 1:1.}
 \label{fig:lya_ha}
\end{figure}

In Fig.\ \ref{fig:lya_ha}, the \lya\ flux is plotted against the \ha\ one for the {\it GALEX} objects . While some objects show a \lya/\ha\ ratio consistent with the case B value, most of them lie well below this line. On the other hand, few objects show a \lya/\ha\ $> 8.7$ at a significant level. The heavily attenuated values can be explained by the high extinction in the UV compared to the optical and to resonant scattering of \lya\ that increases the absorption optical depth. However, the observed dispersion and the points above the case B line could be the result of other parameters that will be discussed below. A convenient way to constrain the dust extinction parameter is to determine the escape fraction of \lya\ as a function of E(B-V).

\begin{figure}[htbp]
 \centering
 \includegraphics[width=9cm,height=7.5cm]{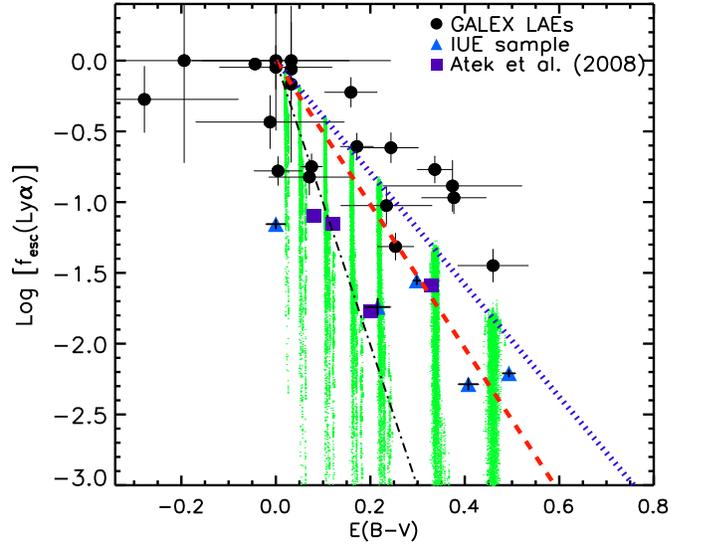}
 \caption{\lya\ escape fraction as a function of dust extinction, observed in $z \sim 0.3$ \lya\ galaxies. The red dashed line represents the best fit to our entire sample ({\it GALEX, IUE}, and Atek et al. samples). 
The dark dot-dashed line represent the best fit determined by \citet{verhamme08} from spectral fitting of $z \sim 3$ LBGs. The blue dotted line corresponds to the escape fraction of the continuum attenuated by dust extinction using the \citet{cardelli89} law. Green points are predictions for \fesc(\lya) using a 3D \lya\ radiation transfer code \citep{verhamme08}. See text for details.}
 \label{fig:fesc_ebv}
\end{figure}

To determine the \lya\ escape fraction we follow  \citet{atek08}. 
The method relies on the fact that \ha\ emission is not prone to complex radiation transport effects but is only affected by dust attenuation. Therefore, correcting the observed \ha\ flux for extinction while assuming a case B recombination theory \citep{osterbrock89}, one can estimate the intrinsic \lya\ flux. The \lya\ escape fraction is then given by:
\begin{equation} 
 f_{esc}(\mathrm{Ly}\alpha) = f(\mathrm{Ly}\alpha)/(8.7 \times f(\mathrm{H}\alpha)_{C}) ,
\end{equation}       
where $f$(\lya) is the observed flux and $f$(\ha)$_{C}$ is the extinction-corrected \ha\ flux.

Figure \ref{fig:fesc_ebv} shows our empirical \fesc\ values as a function of the nebular extinction for the {\it GALEX} 
and {\it IUE} samples described above. \fesc\ values for 4 nearby galaxies from \citet{atek08}, with \ewlya\ $> 0$ \AA, are also shown.
This figure summarizes much information with several implications for \lya\ physics. 

First, we find that \fesc\ is not constant. It spans a wide range of values, typically from \fesc\ $\sim 0.5$ to 100 \%
in the  {\it GALEX} sample.
Second, the \lya\ escape fraction is clearly sensitive to the dust extinction and an anti-correlation is observed. We performed a linear least-squares fit to this anti-correlation. The red dashed curve is the best solution found corresponding to 
\begin{eqnarray} 
\hspace{0.5cm} f_{esc}(\mathrm{Ly}\alpha) &=&  10^{-0.4 \  k(\mathrm{Ly}\alpha)  \ E(B-V)} \ \ ;  \hspace{0.2cm}   \  k(\mathrm{Ly}\alpha) \sim 12.7 \pm 0.4 .
\end{eqnarray}      
 The extinction coefficient at \lya\ wavelength goes from $k(1216)  \sim 9.9$ to $12.8$ for \citet{cardelli89} and \citet{calzetti00} laws, respectively. Here, $k$(\lya) derived from the fit takes into account the averaged effects of all processes affecting the \lya\ escape, such as the resonant scattering experienced by \lya\ photons, which increases their mean path and therefore the effective dust optical depth,
or velocity fields in the gas or the ISM geometry, which may ease the escape of \lya.
The observed scatter around our mean relation is most likely indicative of this multi-parameter process. Using hydrodynamical cosmological simulations of \lya\ emitters, \citet{dayal09} find similar trend between \fesc(\lya) and extinction.
%
For similar extinctions, the \lya\ escape fraction tends to be higher on average  in {\it GALEX} \lya\ galaxies than in local objects. 
In the $z \sim 0$ objects, \fesc\ never exceeds $\sim 10\%$, whereas it covers a wide range of values reaching 100\% at $z \sim 0.3$. 
This difference may be the result of different selection effects. The  {\it GALEX} galaxies are selected from their \lya\ emission amongst spectra taken from a blind search, whereas the local objects are from a sample of specific, optically-selected starburst galaxies.


Third, several objects show \fesc(\lya) greater than the escape fraction expected for the continuum near \lya\ (blue dotted line) as given by \fesc(cont) $= 10^{-0.4\ k(1216) \ E(B-V) }$, where the most favourable value $k(1216) \sim 9.9$ is adopted from the Cardelli et al. law. 
Although \lya/\ha\ ratios exceeding the theoretical value have already been found in local starbursts \citep{atek08},
they only occur locally, in spatially resolved objects, where this easily can be explained by a local \lya\ ``excess''
due to scattering. Here, in some objects, it is the ``global'' \lya\ escape fraction determined from the integrated spectra that is found to 
be higher than expected from the most favourable (i.e.\ flattest) attenuation law.
These objects may be observational evidence for a multi-phase configuration of the ISM \citep{neufeld91,hansen06,finkelstein09a}, where dust is primarily distributed in cold neutral clouds with an ionized inter-cloud medium. By reflecting on the cloud surface, \lya\ photons will be easily transmitted through the ionized medium. Alternatively, Scarlata et al. (2009, submitted) have advocated for a clumpy dust distribution scenario able to reproduce the observed \lya, \ha\ and \hb\ intensities without the need for a two-phase model responsible for different paths for \lya\ and \ha. 
 
Approximately 2/6 of  these objects also show a relatively high \lya\ equivalent width (\ewlya $\sim$ 100-150 \AA), as may
be expected for a clumpy ISM. Furthermore, \fesc(\lya) higher than the UV continuum
could also be due to orientation effects in objects with an aspherical
ISM, e.g. in conical outflows, into which \lya\ would
be ``channeled'' more effectively than continuum radiation.
The \lya\ escape fractions of objects below the attenuation curve of the continuum can be quantitatively reproduced by models using a homogeneous ISM. However, this does not exclude that clumping may also play a role in these objects.





We performed an extensive grid of 3D \lya\ radiation transfer simulations in homogeneous, spherically expanding shells (Hayes et al. in prep), around a central emitting source of UV continuum plus the \lya\ line, using an updated version of the MCLya code \citep{verhamme06}. We overplot in Fig. \ref{fig:fesc_ebv} the predictions of \fesc\ as a function of E(B-V) by using all possible values of the remaining parameters affecting \fesc: the expansion velocity of the neutral gas, \hi\ column density, Doppler parameter $b$ and FWHM(\lya) of the input emission line \citep[for details, see][]{atek09a}. The model E(B-V) is computed from the predicted UV continuum attenuation, assuming the same extinction law as above. The model grid covers the bulk of the observed variations in \fesc\ and E(B-V). The homogeneous, spherical shell models may in principle be able to explain the majority of the objects, although tailored models including all observational constraints are needed to confirm this.  However, the \fesc\ values above the continuum attenuation curve (blue dotted line in Fig.\ \ref{fig:fesc_ebv}) observed in 6 objects cannot be explained with these models, since \lya\ photons cannot be less attenuated than the continuum in a homogeneous ISM.
New radiation transfer computations in clumpy media are underway to examine this interesting behavior.




\section{Discussion}

We have presented here an estimation of the mean \lya\ escape fraction as a function of the extinction and how different parameters can alter this simple relation. We have carefully chosen our aperture (5\arcsec\ slit) in optical spectroscopy in order to obtain Balmer fluxes comparable to \lya\ ones obtained by the {\it GALEX} grism. This allows us to estimate the extinction and the escape fraction for the whole galaxy.


To keep the same consistency between the different samples, we decided to minimize selection effects by retaining only the net \lya\ emitters in  both the {\it IUE} and the imaging samples. We recall that the retained {\it IUE} large aperture (20\arcsec\ $\times$ 10\arcsec) UV observations and ground-based optical spectroscopy of local starbursts are all aperture matched. Furthermore, for \lya\ imaging objects, the {\it HST} aperture is large enough to encompass the entire \lya\ emission region. However, while the slitless mode of {\it GALEX} enables us to recover the diffuse \lya\ emission, this is not necessarily the case for {\it IUE} observations. The size of the large aperture may remain insufficient for some nearby objects in order to encompass the scattered photons across a large area of the galaxy, as usually indicated by the large extent of the \hi\ gas. Therefore, \fesc(\lya) could be slightly underestimated. On the contrary, as mentioned earlier,  the $z \sim 0.2 - 0.3$ objects are selected on the grounds of their \lya\ strength. This will likely favor the high \lya\ escape fractions. One should then keep in mind that these opposite effects contribute to stretch the deviation from our best fit of this compilation of data points. 
Similar \lya\ equivalent width criteria are commonly applied to select high-z LAEs, implying most likely relatively high escape fractions and a less severe discrepancy between \lya\ and non-resonant radiation for this class of objects. 
A blind search of \lya\ emitters would then find lower \fesc(\lya) than the {\it GALEX} objects. This is already found in our current double 
blind survey in \lya\ and \ha\ at $z \sim 2.2$, where an average escape fraction of $\sim 4.5$ \% is measured (Hayes et al. in prep).


%

In contrast with other emission lines, the dust extinction is only one of several parameters governing \fesc(\lya), and the extent of the dispersion around the fit is a good illustration. Given the importance of an accurate estimation of \fesc(\lya), one needs to quantify the kinematical effects by measuring the cold ISM velocity in these objects; an additional step toward a precise calibration of \fesc(\lya). A more detailed investigation of the physical properties and SED modeling of our sample will be carried out in subsequent publications (Atek et al. in prep).
\begin{acknowledgements}
We thank Daniela Calzetti, who kindly put her UV-optical spectra of the IUE sample at our disposal. The work of DS and MH is supported by the Swiss National Science Foundation. HA and DK are supported by the Centre National d'Etudes Spatiales (CNES). G\"O is Royal Swedish Academy of Sciences Research Fellow supported by a grant from the Knut and Alice Wallenberg Foundation. G\"O acknowledges support from the Swedish research council. JMMH is partially funded by Spanish MICINN grant AYA2007-67965. This work is based on observations made with ESO Telescopes at La Silla Observatories under programme ID 082.B-0392. 
\end{acknowledgements}
\bibliographystyle{aa}
\bibliography{references}

\end{document}